# A scalable dense silicon qubit architecture with logical operations

Masahiro Tadokoro[1] and Tetsuo Kodera[1]

[1] *Institute of Science Tokyo, Japan*

Fault-tolerant quantum computing (FTQC) requires a large number of physical qubits to encode logical qubits with sufficiently low error rates. Silicon spin qubits are a promising platform because of their small footprint, compatibility with semiconductor manufacturing technologies, and demonstrated high-fidelity operations. However, realizing a scalable architecture that supports quantum error correction, qubit transport, wiring, and logical gate operations remains a major challenge. Here, we propose a scalable dense silicon spin qubit architecture designed to support large-scale FTQC. The architecture consists of qubit memories, data qubit lanes, and cross junctions. Qubit memories store logical qubits encoded using the rotated surface code and support logical single-qubit operations. Data qubit lanes enable high-speed one-dimensional spin shuttling for connectivity between logical qubits, while cross junctions provide routing functionality and implement transversal logical CNOT gates. Using experimentally demonstrated silicon spin qubit parameters, we evaluate logical error rates as a function of code distance and estimate that a code distance of at least 21 is required to achieve logical error rates below $10^{-10}$ per syndrome cycle. We further analyze wiring constraints based on CMOS-compatible multiplexed control schemes and show that, with optimized wiring efficiency, more than 120,000 logical qubits could be integrated on a 100 mm × 100 mm chip. These results provide a practical roadmap toward large-scale FTQC based on silicon spin qubit technology.

## I. INTRODUCTION

The primary challenge for large-scale fault-tolerant quantum computation (FTQC) is the requirement for a vast number of physical qubits to encode a single logical qubit with high reliability. In this regard, silicon spin qubit systems are promising because they have a small physical qubit footprint and possess gate fidelities necessary for fault tolerance [1] -[13]. Silicon spin qubit architectures that leverage these advantages have been proposed to date. For example, an architecture that optimizes wiring by connecting a common wiring line to multiple qubits [14] and architectures that maintain uniformity in qubit characteristics by sparsely distributing qubits [15][16] have been proposed. While these proposals aim to address challenges in device fabrication, they have limitations when considering the implementation of large-scale FTQC. In architectures with optimized wiring, independent qubit operations are not possible, resulting in slower qubit operations. Consequently, an unnecessarily large code distance must be adopted, which may prevent the preparation of a sufficient number of logical qubits. Furthermore, if qubits are sparsely distributed, not only does the qubit footprint increase, but shuttling also takes time, requiring a large code distance. Therefore, these conventional proposals have not been discussed from the perspective of implementing logical operations, and it remains unclear whether they can support large-scale FTQC.

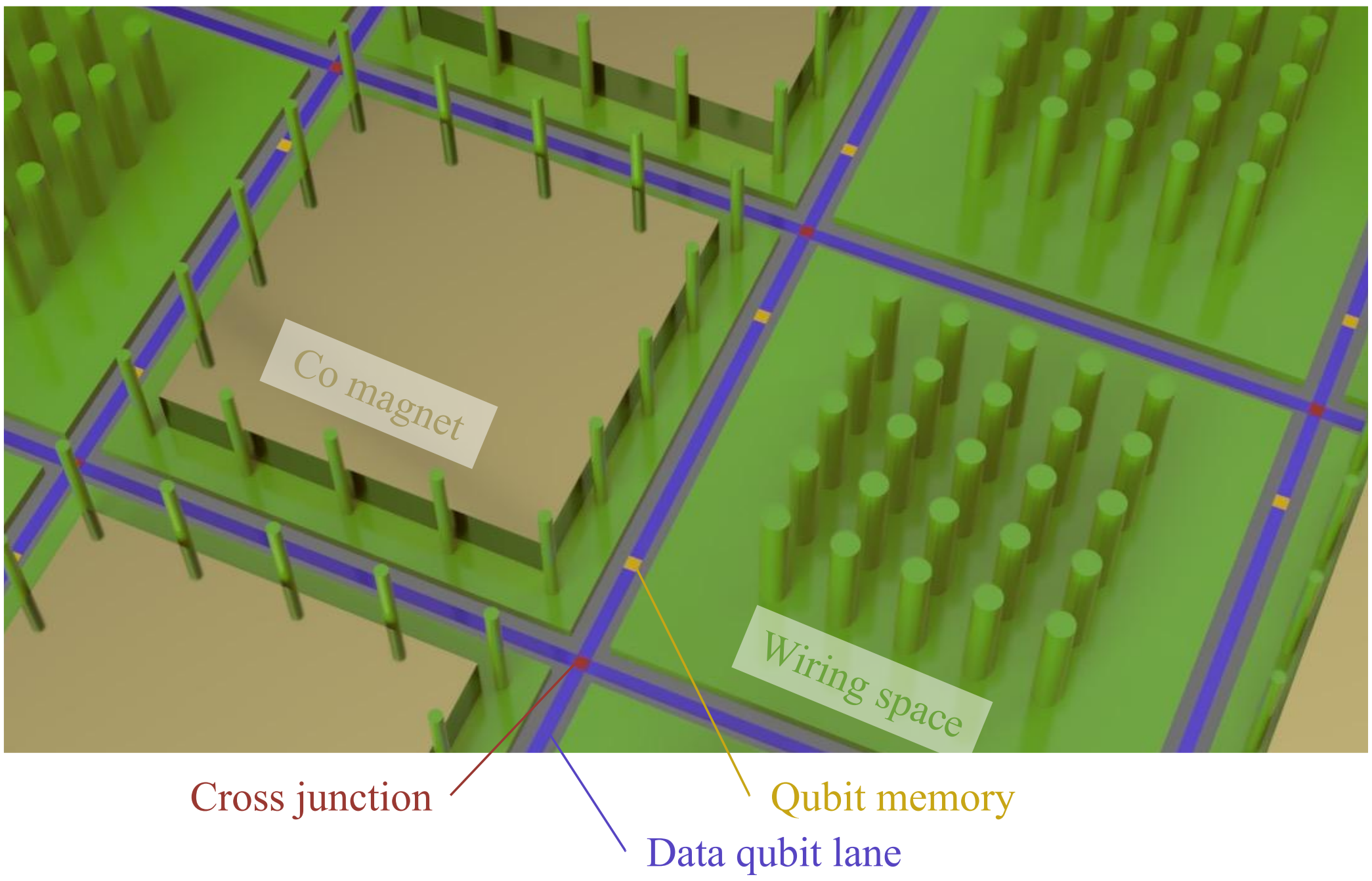

Figure 1 An image of the entire architecture. The yellow squares arranged around the Co magnet represent qubit memories, and independent qubit operations are possible within each qubit memory. The data qubit lane, shown in purple, connects the qubit memories. The cross junctions where the data qubit lanes intersect are represented by red squares. Control is achieved via wiring connected to the adjacent green wiring space.

This paper proposes a scalable architecture that consists of dense silicon qubit arrays to implement logical operations. This architecture is based on a dense silicon qubit array capable of implementing a quantum error correction code with the necessary code distance based on the gate fidelity and gate times obtained from existing silicon qubit systems. While the feasibility of high-density wiring around qubits in a dense silicon qubit array poses a challenge, this basic structure allows the integration of readout resonant circuits, which are a bottleneck for scaling, using a feasible number of gate stacks.

Furthermore, by repeating this basic structure in a two-dimensional array, it is possible to integrate 120,000 logical qubits into a 100 mm × 100 mm chip under the assumption of maximum wiring efficiency.

This paper presents a pathway to large-scale FTQC that leverages the advantages of silicon qubit systems: a small physical qubit footprint and high gate fidelities.

## II. ARCHITECTURE

Figure 1 provides an overview of the entire architecture. A “qubit memory” is arranged around the Co magnet, consisting of a qubit array capable of independent qubit operations, and holds logical qubits [17]. A “data qubit lane” capable of shuttling is positioned between adjacent qubit memories, allowing only the data qubits of the logical qubits to be shuttled. The “cross junction,” where the vertical and horizontal data qubit lanes intersect, enables shuttling in both vertical and horizontal directions and functions as an “intersection” of the data qubit lanes. Furthermore, it can store the data qubits of two logical qubits, enabling the implementation of a transverse entangling gate [18]. Co magnets are sparsely arranged, and the surrounding space serves as wiring space. The wiring for the qubit memory, data qubit lane, and cross junction is routed to the nearest wiring space, where it connects to external electronic circuits via vertical vias. The following sections describe the details of each component included in this architecture.

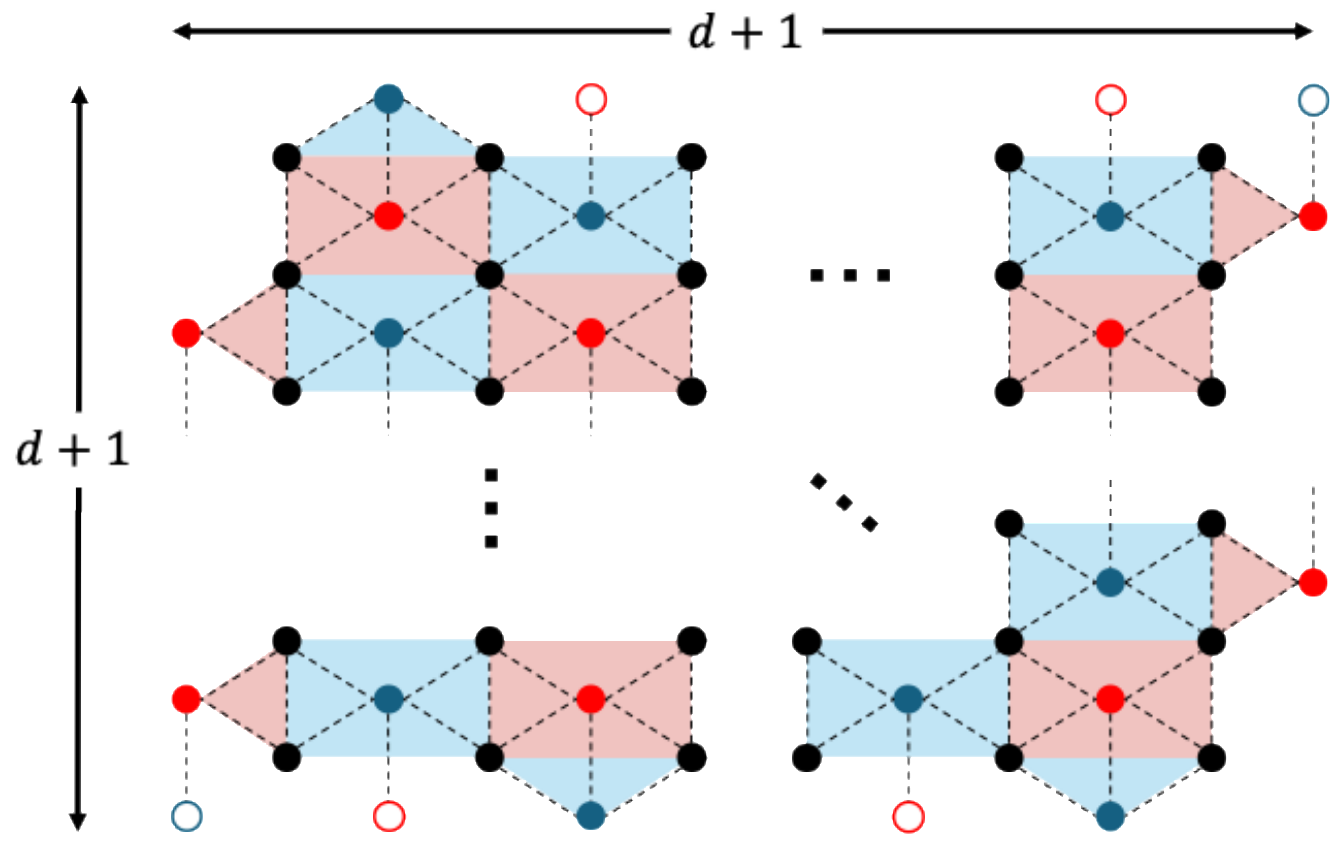

Figure 2 Schematic of the qubit array in a qubit memory. The black circles represent data qubits, the blue circles represent X ancilla qubits, and the red circles represent Z ancilla qubits. The dotted lines represent qubit connectivity. The ancilla qubits shown as open circles do not contribute to encoding but are necessary for parity measurements.

### A. Qubit memory

Figure 2 shows a schematic of the qubit array in a qubit memory that constitutes the logical qubit encoded by the rotated surface code. Among various candidates for QEC codes, the rotated surface code is the leading code due to its high error threshold and the requirement of only local, nearest-neighbor connectivity [19]. Although the overhead associated with the relatively low encoding rate remains a major bottleneck for this code, as shown in the simulation results presented below, silicon spin qubits have a small footprint and can ensure the necessary code distance.

The basic unit is the triangular triple QD [20]; black circles represent data qubits, blue circles represent X ancilla qubits, and red circles represent Z ancilla qubits, while the dotted lines represent qubit connectivity. When encoding a rotated surface code with code distances $d$, a $(d+1)\times(d+1)$ qubit array is required, considering the ancilla qubits at both ends of the code. Let the inter-QD distance be $d_{QD}$; the qubit array dimensions are then $d_{QD}\times 2\sqrt{3}\times d$ in width and $d_{QD}\times d$ in height. Since the architecture discussed in [17] can maintain independent qubit operations for a QD array up to 5 μm × 5 μm, a single qubit memory can store up to 16 logical qubits assuming $d=21$ and $d_{QD}=50\text{ nm}$ for instance.

However, a key challenge here is routing the connections to the wiring space. There have been no previous studies implementing 3D wiring for large-scale qubit arrays. Therefore, as shown in Figure 1, the wiring to the vast wiring space, which can utilize vertical vias, must be stacked on a two-dimensional plane. Ideally, we would store multiple logical qubits in a large-scale qubit memory and perform operations within it. However, achieving this would require stacking dozens of metal gates, making fabrication difficult. Hence, we will limit the number of logical qubits stored in a single qubit memory to one and reduce the size of the qubit memory. In this situation, 5 μm × 5 μm qubit memory can implement $d=85$ with $d_{QD}=50\text{ nm}$.

Figure 3 shows a schematic diagram of the wiring in a qubit memory. To optimize the wiring, we assume that only a plunger gate is fabricated above each QD, and that exchange interactions between spins are controlled solely by the voltage applied to the plunger gates [21]. In addition, to apply the gradient magnetic field required for EDSR implementation, the lowest gate layer, which is closest to the QD, is fabricated with Co and magnetized [17]. By utilizing this magnetized Co gate and the Co magnet adjacent to the qubit memory, each data qubit can be operated independently, enabling the efficient implementation of logical Pauli gates and logical Hadamard gates. In this case, $(d+1)/2+1$ layers of metal gates are required for the wiring connections. Furthermore, since the conversion factor between each plunger gate and each QD is expected to be large, we assume the introduction of a virtual gate [22].

### B. Data qubit lane

Figure 4 shows a schematic diagram of a data qubit lane. Within the data qubit lane, spins are shuttled in a one-dimensional direction while maintaining the array of data qubits, enabling high-speed, high-fidelity conveyor-mode shuttling [23]. The gray area represents a region where the SOI layer has been etched. By restricting the path to one dimension, spins can shuttle without tunneling to other columns, resulting in a structure that does not require a confinement gate. A metal gate is fabricated to apply control signals to this path, and shuttling is performed by applying a wave potential that propagates along the path.

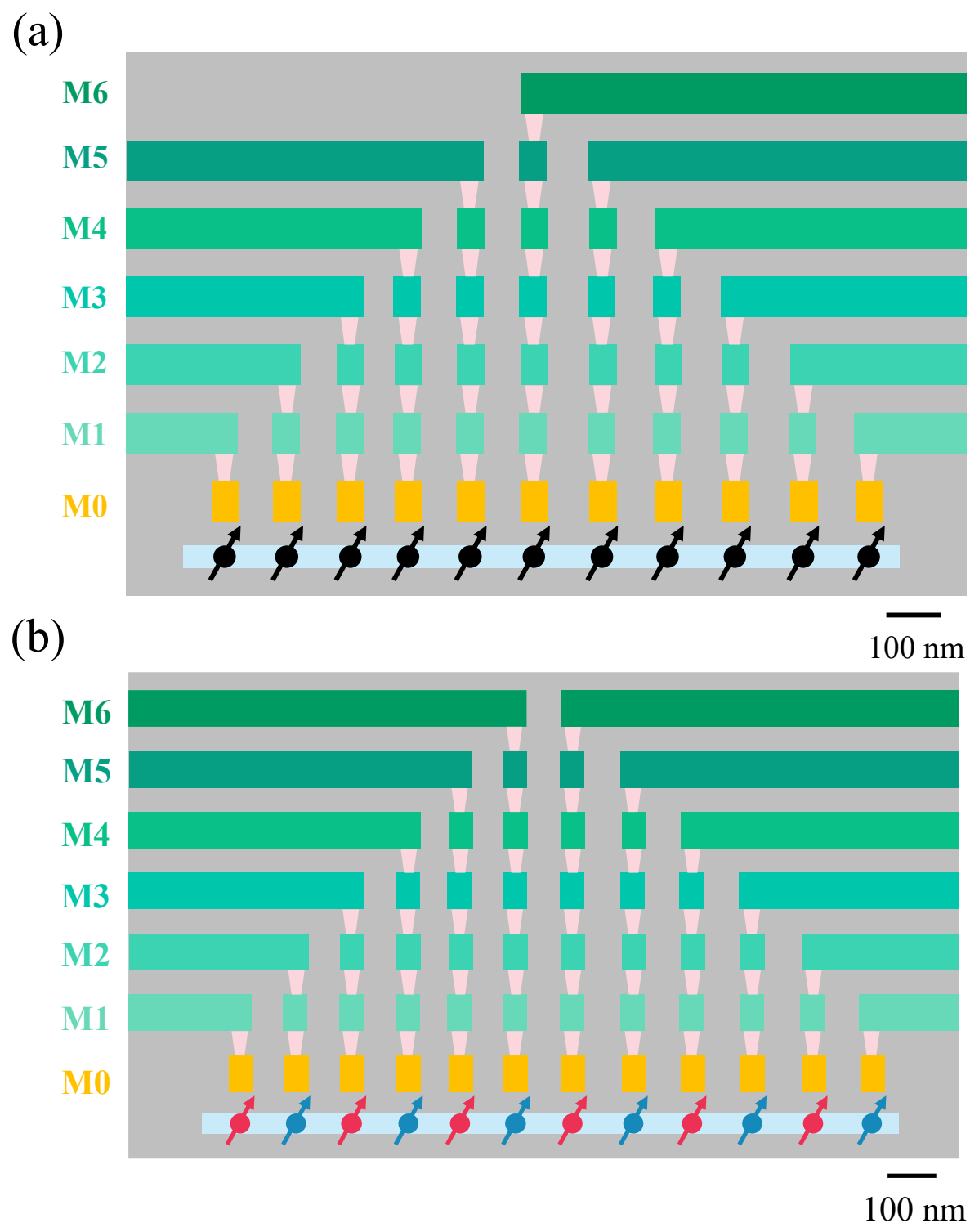

Figure 3 Schematic diagram showing the wiring for (a) data qubits and (b) ancilla qubits in a qubit memory with $\boldsymbol{d} = \mathbf{11}$. Si in which the QDs are formed is shown in light blue, and this is where spins are stored. The surrounding gray regions are insulators. A magnetized Co gate is fabricated above each QD for spin operations, and the wiring is connected by a total of seven layers of gate stacks (greens) through vias (pink). These wires are connected to the left and right wiring spaces of the qubit memory.

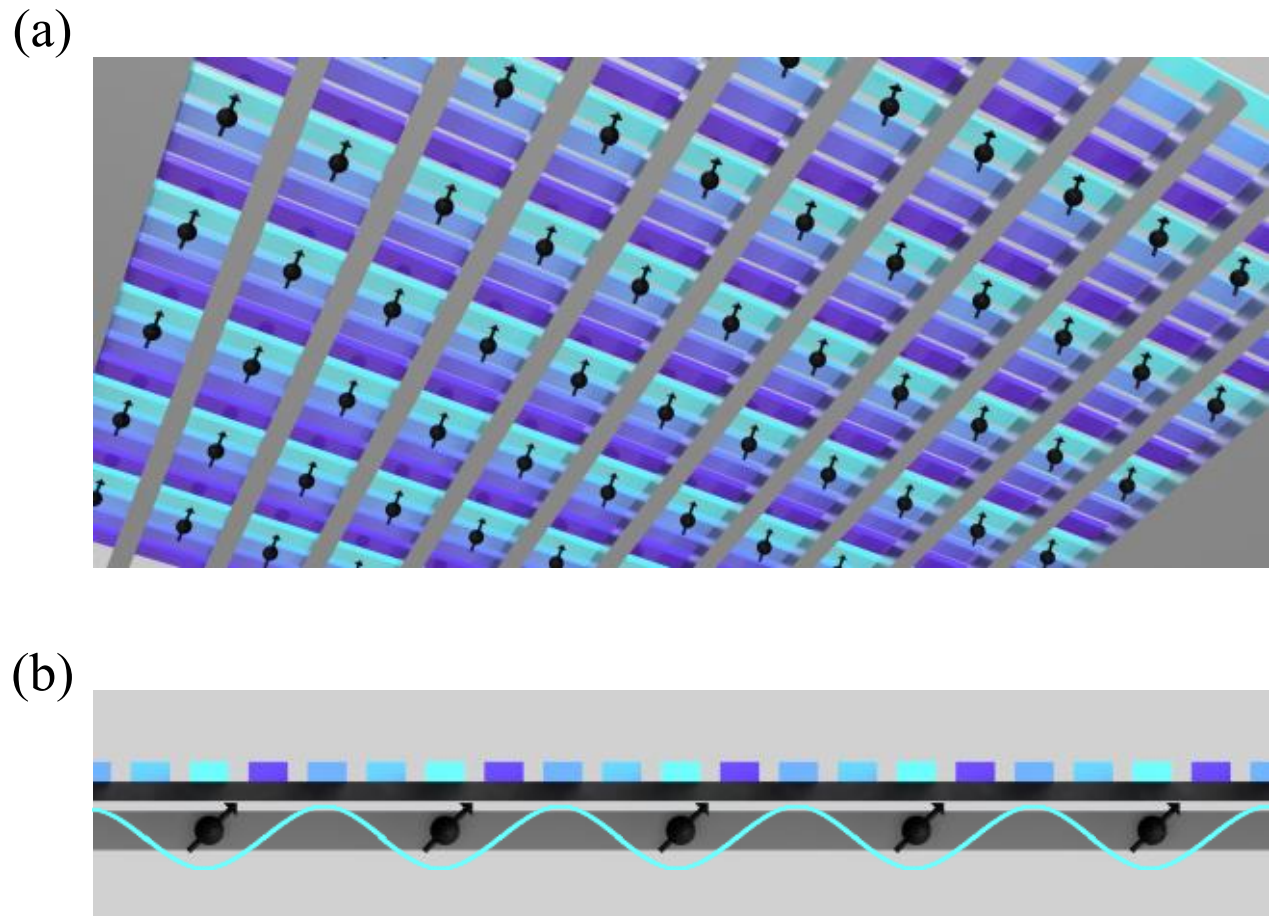

Figure 4 (a) Schematic diagram of the data qubit lane. (b) Schematic diagram of the wiring in the data qubit lane. This is a cross-sectional view along the direction in which the spin shuttles. This figure illustrates an example of shuttling using four sinusoidal control signals, as described in [24], with the gate voltages applied to the gate wiring indicated by different colors.

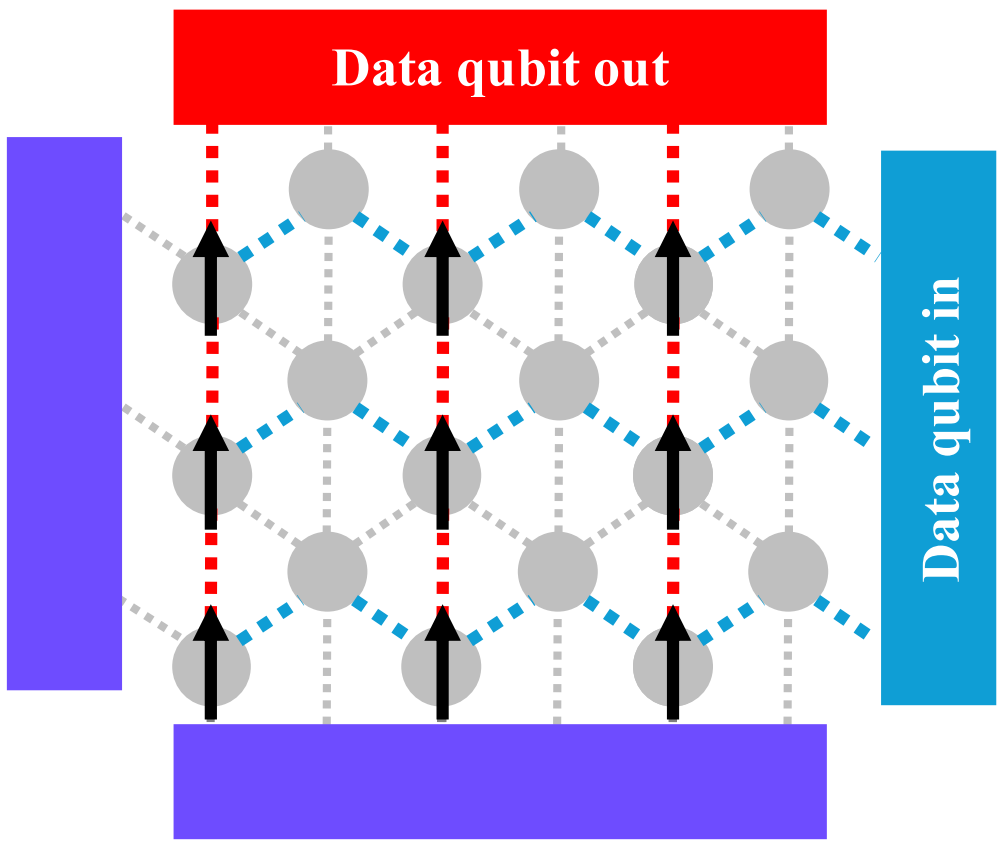

Figure 5 The flow at the cross junction where a data qubit shuttled from the right data qubit lane to the upper data qubit lane is directed. For simplicity, we show the case where $d = 3$. Black arrows represent data qubits, gray circles represent QDs, and surrounding squares represent data qubit lanes. Data qubits flow in via the blue path from the right data qubit lane into the cross junction and then flow out via the red path to the upper data qubit lane. This allows data qubits to be moved two-dimensionally within the architecture.

### C. Cross junction

The cross junction, where the vertical and horizontal data qubit lanes intersect, is also constructed using the QD array like qubit memory. However, its function differs from that of qubit memory, as it serves two purposes.

The first purpose is to act as an "intersection" for the data qubit lanes, aligning the direction in which data qubits are shuttled. Unlike the data qubit lane, the cross junction requires data qubits to be shuttled in both the vertical and horizontal directions. Therefore, bucket-brigade mode shuttling is employed [22]. Figure 5 illustrates the flow in which data qubits shuttled from the right data qubit lane are directed to the upper data qubit lane at the cross junction.

The second purpose is implementing logical CNOT gate operations. As representative methods for implementing logical CNOT gates in rotated surface codes, there are two approaches: Transversal entangling gates and Lattice Surgery [18]. In this architecture, to avoid implementing a resonant circuit for spin readout, we employ Transversal entangling gates that do not require qubit readout. They are implemented by simultaneously applying physical CNOT gates individually to all corresponding data qubit pairs on both the control and target logical qubits. Since the cross junction does not execute QEC cycles and no ancilla qubits exist, there are QDs capable of storing two data qubits. Therefore, it is possible to shuttle the two data qubits to the cross junction for storage, thereby implementing a transversal CNOT gate (Figure 6).

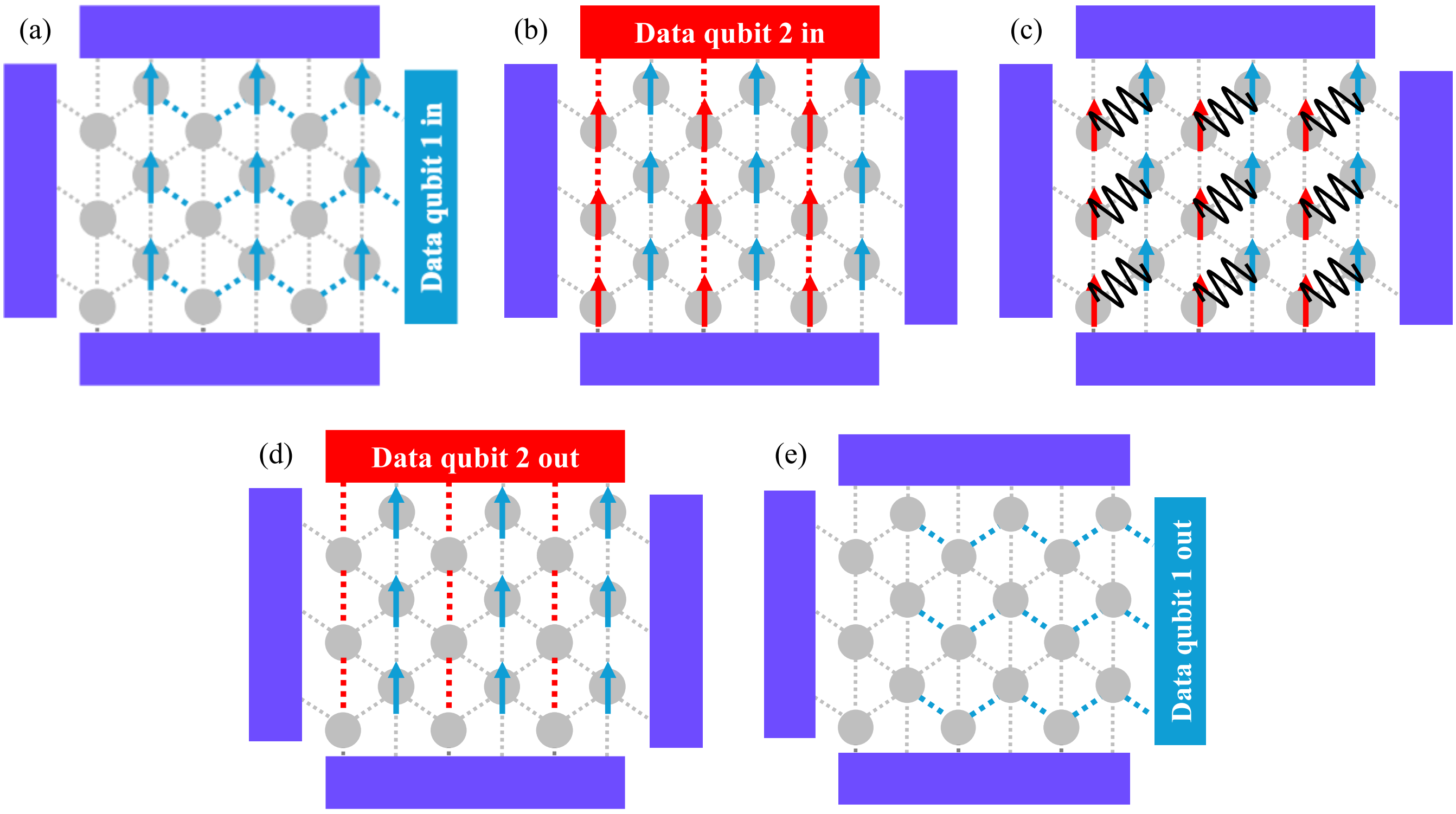

Figure 6 The flow for implementing a transversal CNOT gate between data qubits shuttled from the right (Data qubit 1) and upper data qubit lanes (Data qubit 2) to the cross junction. First, Data qubit 1 flows in from the right data qubit lane to the cross junction (a). Next, Data qubit 2 flows in from the upper data qubit lane (b). Then, a transversal CNOT gate is implemented between the respective physical qubits (c). Finally, the data qubits flow out in the reverse order of their inflow (d), (e).

## III. DISCUSSIONS

Here, we discuss the performance characteristics and scalability of the proposed architecture. We evaluate the variation in logical errors due to code distance, which determines qubit memory size, by considering the specific steps of the syndrome cycle. We also discuss the variation in the number of logical qubits that can be integrated onto a single chip, taking into account practical wiring integration requirements.

### A. Logical error rates

Firstly, we evaluate logical error rates expected on this architecture. Although precise estimation of logical errors requires precise estimation of the physical noise channel, we employ the method described in [25], which estimates logical error rates on quantum memory experiment using general qubit characteristics. Table 1 summarizes the silicon spin qubit characteristics that have been experimentally evaluated to date. Note that this contains only the best results for each characteristic.

Table 1 Qubit configuration parameters assumed.

| Quantity | Description | parameter | reference |
|---|---|---|---|
| $T_1$ | Qubit relaxation time | 1 s | [26] |
| $T_2$ | Qubit dephasing time | 100 μs | [27] |
| $p_{1Q}$ | Single-qubit gate error probability | $10^{-4}$ | [28] |
| $p_{2Q}$ | CNOT gate error probability | $10^{-3}$ | [29][30] |
| $p_{\text{meas}}$ | Measurement error probability | $10^{-3}$ | [26] |
| $p_{\text{init}}$ | Qubit initialization error probability | $10^{-3}$ | [26] |
| $t_{1Q}$ | Single-qubit gate time | 10 ns | [31] |
| $t_{2Q}$ | CNOT gate time | 10 ns | [32] |
| $t_{\text{meas}}$ | Measurement time | 1 μs | [33] |
| $t_{\text{init}}$ | Qubit initialization time | 1 μs | [33] |

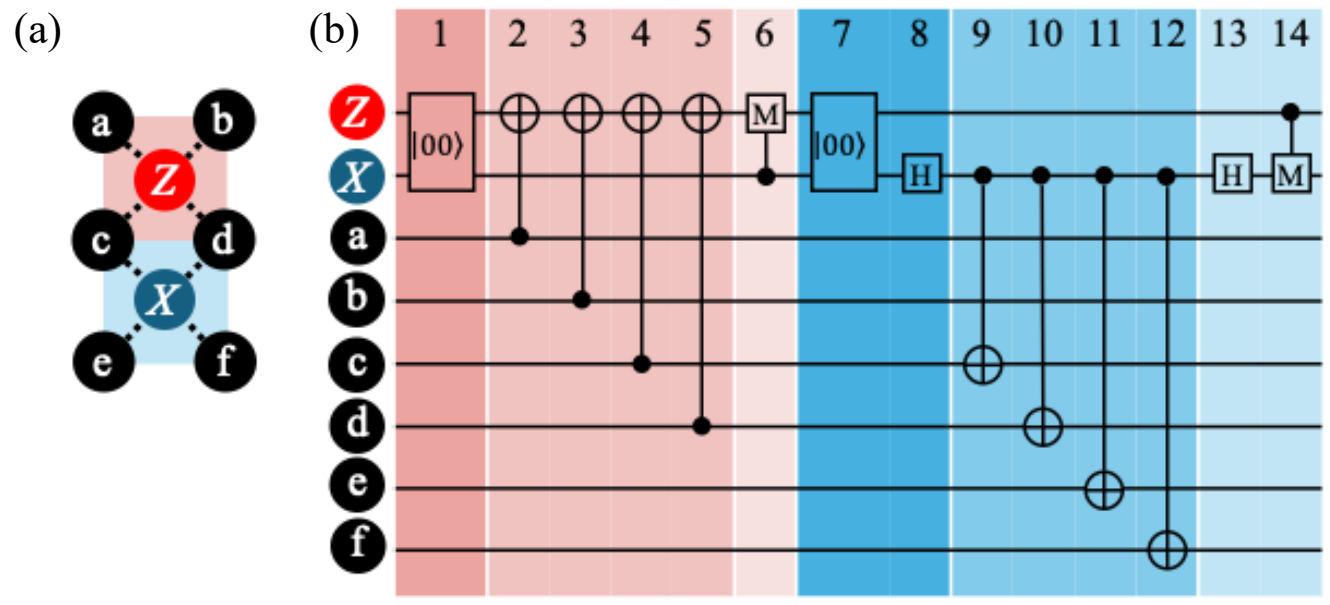

Figure 7 A schematic diagram of a syndrome cycle on a rotated surface code. (a) A physical qubit connectivity. A center red (blue) qubit is a Z (X) ancilla qubit, and other black qubits are data qubits. (b) A circuit diagram of a syndrome cycle. The red (blue) region corresponds to Z (X) syndrome measurement. Timestep 1 (Timestep 7,8) contains state preparation, Timestep 2-5 (Timestep 9-12) contains four consecutive CNOT operations, and Timestep 6 (Timestep 13,14) highlights the measurements on Z (X) syndrome measurement.

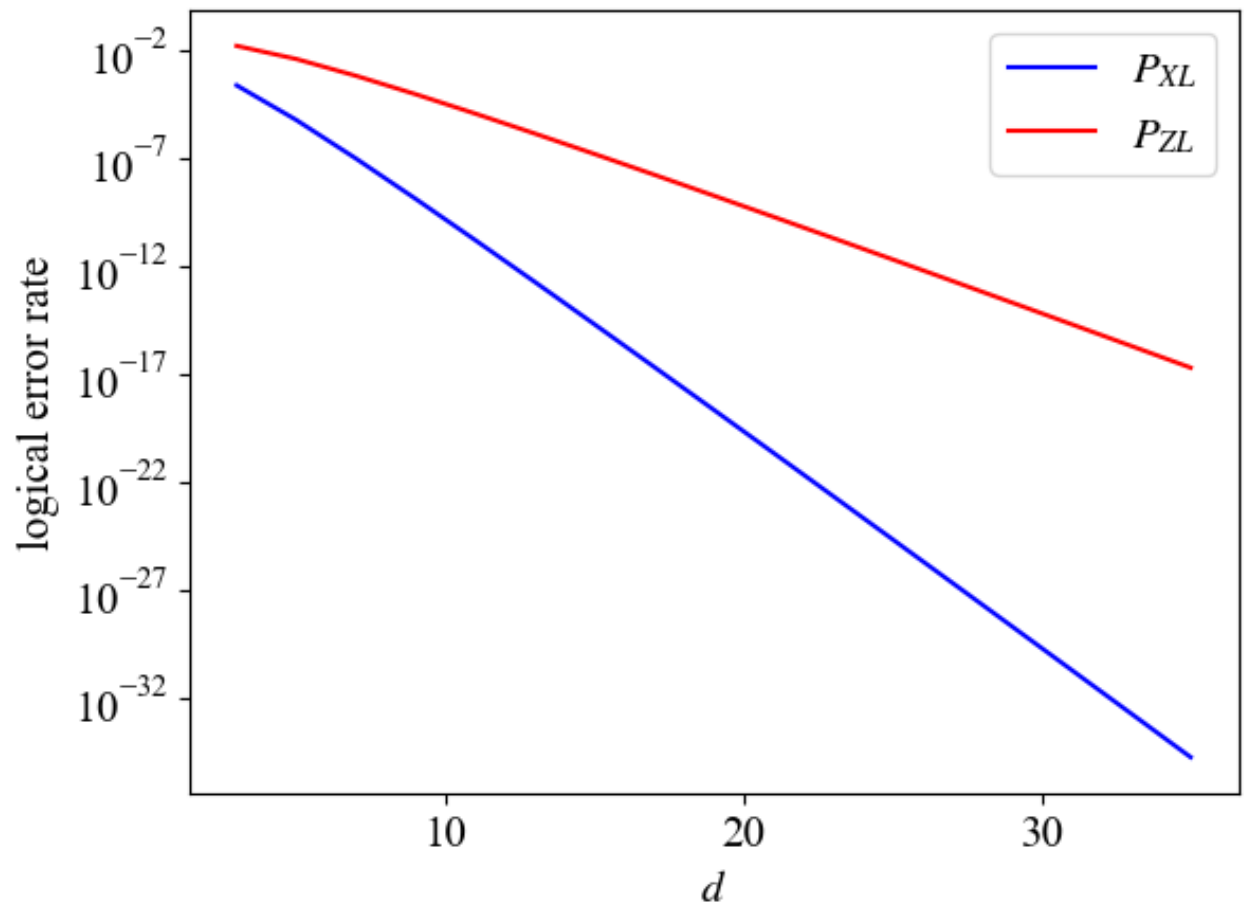

Figure 8 The simulation results of $\boldsymbol{P_{XL}}$ and $\boldsymbol{P_{ZL}}$ per syndrome cycle as a function of $\boldsymbol{d}$.

For defining syndrome cycle time ($t_{\mathrm{cycle}}$), we assume physical quantum circuit of the syndrome cycle as Figure 7. Here, we assume that a PSB is used to measure the ancilla qubit. In this case, we assume that the X (Z) ancilla qubit is used to read out the Z (X) ancilla qubit in a Z (X) syndrome measurement, and that parity measurements are performed on the two ancilla qubits [34]. Since the two syndrome measurements cannot be executed in parallel, the syndrome cycle is constructed using a circuit as shown in Figure 7 (b). The cycle time $t_{\mathrm{cycle}}$ in this case is expressed as follows:

$$t_{\mathrm{cycle}} = 2t_{\mathrm{init}} + 2t_{1\mathrm{Q}} + 8t_{2\mathrm{Q}} + 2t_{\mathrm{meas}} \qquad (1)$$

Figure 8 shows the result of simulating X logical error rates ($P_{XL}$) and Z logical error rates ($P_{ZL}$) per syndrome cycle as a function of $d$. Since $T_1 \gg T_2$, $P_{ZL}$ is always larger than $P_{XL}$. To obtain both $P_{XL}$ and $P_{ZL}$ less than $10^{-10}$ for instance, it can be estimated that a code distance of $d \geq 21$ is required.

This result assumes that decoding is performed by a Minimum Weight Perfect Matching (MWPM) decoder and does not reflect non-Markovian noise such as 1/f noise or spatially correlated noise such as crosstalk. Furthermore, the unitary control error is modeled as a stochastic Pauli noise channel using the Pauli twirl approximation [25]. Moreover, this result only evaluates quantum memory performance and does not include errors associated with shuttling, cross-junction operations, or logical gate execution.

Therefore, this simulation result indicates the minimum required size of a physical qubit array for silicon spin qubits. However, as discussed above, the code distance that can be implemented using a single qubit memory is sufficient for the required code distance, and it is possible to increase the code distance as needed. Furthermore, by applying codes that are better suited to the noise channel, such as the XZZX code [35], it is possible to reduce the code distance required to achieve the desired logical error rate.

### B. Wirings and Scalability

Secondly, we discuss the interconnection methods that pose challenges in dense qubit arrays and their impact on scalability.

For wiring to qubit memories, we utilize the method proposed in [36] to optimize wiring in the wiring space. Figure 9 shows a circuit diagram of the wiring to the ancilla qubits within the qubit memory. The wiring for the gate stacks drawn from the qubit memory to the wiring space is incorporated into a small number of common data lines. These data lines incorporate transistors controlled by word lines (selecting the row of target physical qubits) and bit lines (selecting the column), enabling the addressing to individual physical qubits. Furthermore, capacitors are connected as floating memory near each physical qubit, allowing the voltage necessary to maintain the physical qubit to be applied even if the connection to the data line is cut by the word/bit line. This structure is identical to conventional dynamic random-access memory (DRAM) and can be fabricated using mature CMOS processes.

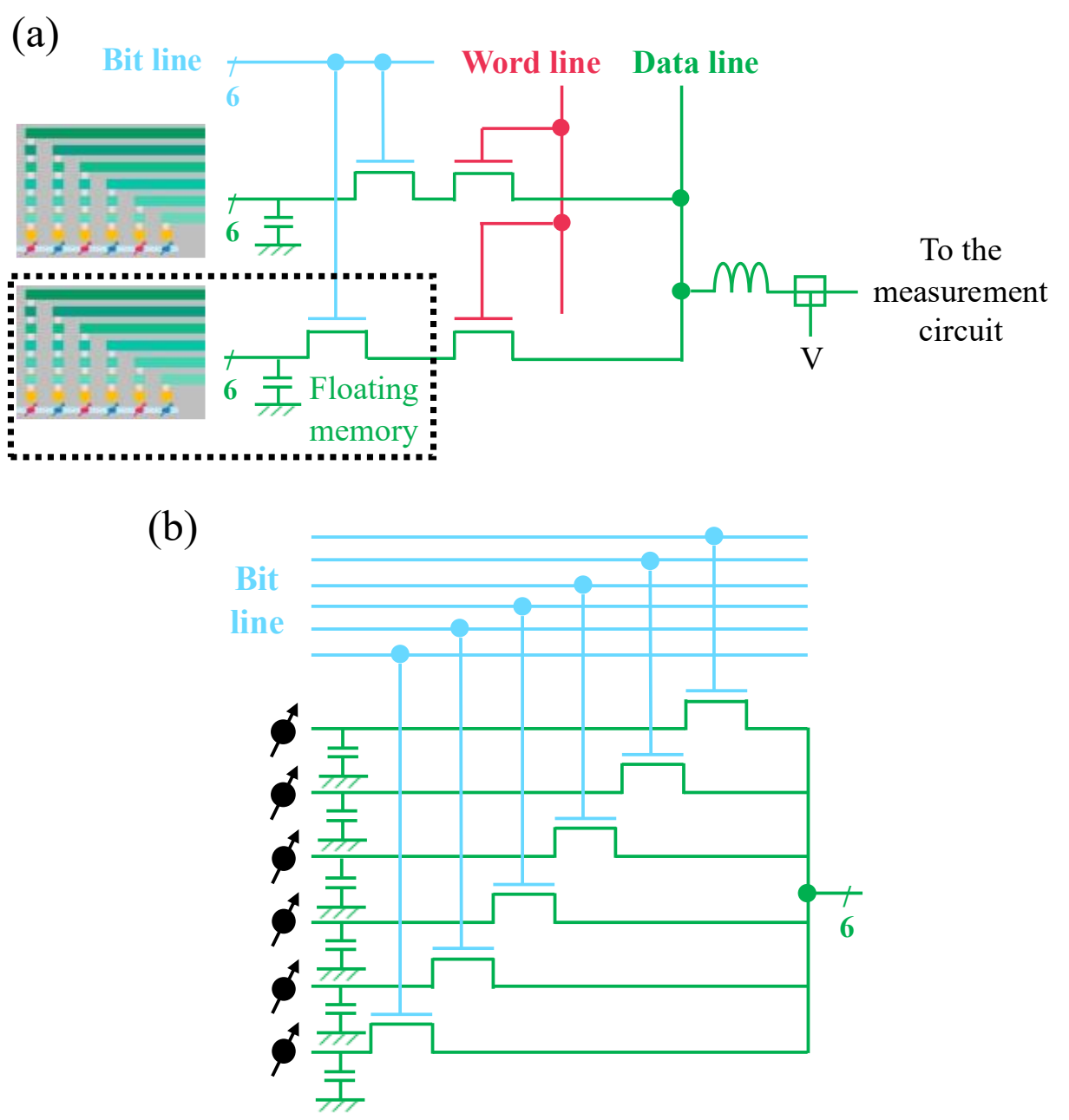

Figure 9 (a) Schematic diagram of the wiring to the ancilla qubits and (b) a detailed diagram within the black dotted line in (a). By connecting a single data line via word/bit lines to the multiple physical qubits that form the qubit memory, it is possible to optimize the wiring while ensuring addressability to individual physical qubits. The data line connected to the ancilla qubits leads to the measurement circuit via a resonant circuit and a bias-tee, enabling spin readout.

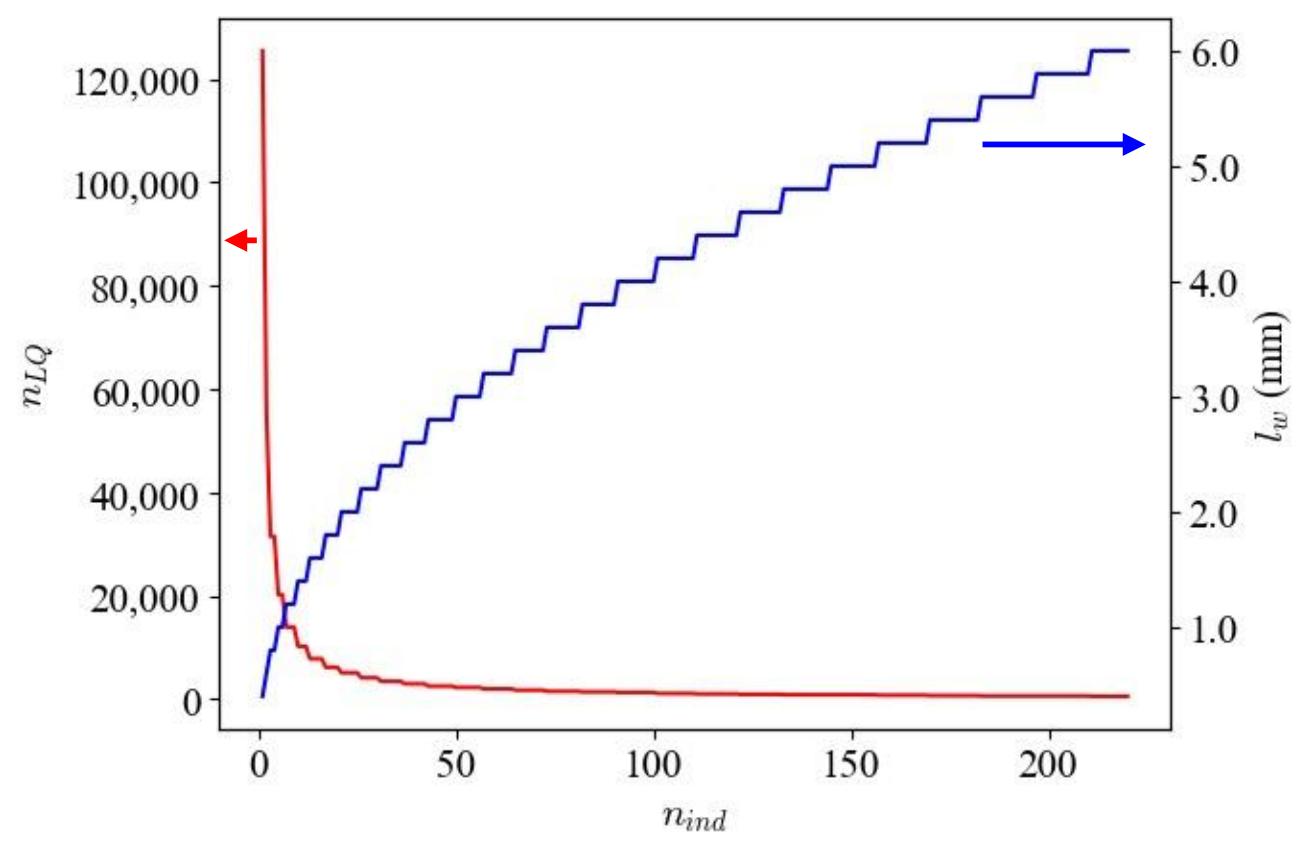

Figure 10 Characterization of the side length of the wiring space ($\boldsymbol{l_w}$) and the number of logical qubits that can be integrated onto a single chip ($\boldsymbol{n_{LQ}}$) as the number of inductors prepared for a single qubit memory ($\boldsymbol{n_{ind}}$) is varied. Since we assume that 0.2 mm × 0.2 mm inductors are embedded in a square wiring space, the result shows a step function.

The data line is connected not only to voltage sources for controlling the potential of each QD but also to resonant circuits for spin readout and a microwave source for MM-ESR. LC circuit consisting of an inductor and the parasitic capacitance of the circuit is used for the resonant circuit. However, this posed a problem because the inductor had to be placed near the qubit, hindering wiring scaling. In contrast, this wiring scheme allows the number of inductors to be reduced by sharing wiring among multiple ancilla qubits. Note that while it is necessary to be able to measure data qubits for logical qubit readout, the data qubit can be measured via quantum non-demolition readout [7] performed by an adjacent ancilla qubit. Therefore, the inductor need only be connected to the ancilla qubit. Instead, the microwave source for MM-ESR is connected to the data qubit to enable single-qubit gate implementation.

The following will discuss how many logical qubits can be integrated using this wiring method. Here, we assume $d = 21$, which yields a logical error rate of $10^{-10}$, and consider integrating LC resonant circuits into the wiring space only for the Z ancilla qubit. In assuming that the inductor size $l_{ind} \times l_{ind}$ is the dominant factor in the wiring space, since it is necessary to integrate the inductors connecting to the four surrounding qubit memories into a single wiring space, the side length of the wiring space ($l_w$) is expressed as follows with the number of inductors prepared for a single qubit memory ($n_{ind}$).

$$l_w = l_{ind} \times \left\lceil \sqrt{4 \times n_{ind}} \right\rceil \text{ (mm)} \tag{2}$$

Assuming that qubit memory is integrated two-dimensionally in a square lattice of length $l_w$, the number of logical qubits that can be integrated onto a single chip ($n_{LQ}$) with a side length of $l_{chip}$ is expressed as follows.

$$n_{LQ} = 2 \times \left\lfloor \frac{l_{chip}}{l_w} \right\rfloor \times \left\lfloor \frac{l_{chip}}{l_w} + 1 \right\rfloor \tag{3}$$

Figure 10 shows plots of $l_w$ (blue line) and $n_{LQ}$ (red line) when $n_{ind}$ is varied for $l_{ind} = 0.2$ mm [37] and $l_{chip} = 100$ mm. By optimizing the wiring with reducing $n_{ind}$, the number of inductors that need to be integrated into the wiring space can be reduced, thereby allowing $l_w$ to be shortened. Consequently, $n_{LQ}$ that can be integrated onto a single chip increases. In wiring one data line to all Z ancilla qubits in the qubit memory, it is possible to integrate more than 120,000 logical qubits. Moreover, the introduction of superinductors may enable an even greater $n_{LQ}$ [38]. Furthermore, reducing $l_w$ is expected to decrease the shuttling time required for implementing logical CNOT gates.

On the other hand, it is important to note that as the number of physical qubits connected to a single data line increases, the number of qubit operations that can be implemented simultaneously decreases. Therefore, since the number of physical qubits in an idle state increases during the syndrome cycle, it is expected that physical errors are more likely to occur during this period. The resulting logical errors can be attributed to the non-uniformity of the physical qubit error rate and can therefore be mitigated through decoder design [39]-[41]. However, considering the effect on logical errors, the number of physical qubits connected to a single data line should be adjusted appropriately based on the size of the wiring space and the desired logical error rate.

Although the proposed architecture provides a scalable framework for qubit arrays and wiring structures, realizing a full stack FTQC system will require careful consideration of the control electronics interfacing with the wiring network. The placement, density, and integration strategy of these electronics may impose additional constraints on scalability and system performance. Future work should therefore investigate the co-design of the qubit array, interconnects, and control hardware.

## IV. CONCLUSIONS

Here, we present a pathway for large-scale FTQC with dense silicon spin qubits consisting of qubit memories, data qubit lanes, and cross junctions. Since this architecture employs the rotated surface code with a high error threshold and straightforward implementation, we can freely set the code distance. Qubit memories not only preserve logical qubits but can also implement logical single-qubit gates. Data qubit lanes support fast data qubit shuttling in one dimension for logical qubit connectivity. Cross junctions enable data qubit handover between data qubit lanes and logical CNOT gate implementations.

Furthermore, we estimate the logical error rate that this architecture is expected to achieve when the code distance is varied. We also evaluate its scalability while taking specific wiring into account. Consequently, this architecture requires a code distance of $d \geq 21$ to achieve a logic error rate of less than $10^{-10}$, and it enables the integration of more than 120,000 logical qubits into a 100 mm × 100 mm chip under the assumption of maximum wiring efficiency.

This architecture fully leverages the strengths of silicon spin qubits: high integration density utilizing mature microfabrication technology, as well as high-speed and high-fidelity operations. These results provide a concrete path toward large-scale FTQC based on silicon spin qubits.

## ACKNOWLEDGEMENT

This research was financially supported by JST Moonshot R&D grant number JPMJMS256H, MEXT Quantum Leap Flagship Program (MEXT QLEAP) grant number JPMXS0118069228, JST ASPIRE grant number JPMJAP25A2, and Grants-in-Aid for Scientific Research grant numbers JP25K24574 and JP23K17327. The authors thank Takeru Sakai for helpful discussions and critical reading of the manuscript.